\documentclass[11pt]{article}
\usepackage{bbm}
\usepackage{amsfonts}
\usepackage{mathrsfs}

\usepackage{amsmath,amssymb}

\usepackage{epsfig}
\usepackage{graphicx}               
\usepackage{url}
\usepackage{hyperref}

\usepackage{pstricks}
\usepackage{color}
\usepackage{cite}

 \newcommand{\prd}{{Phys.~Rev.~D}}
 \newcommand{\apjl}{{Astrophys.~J.~Lett.}}
 
 \newcommand{\apj}{{Astrophys.~J.}}

\newcommand{\nc}{\newcommand}

\nc{\beq}{\begin{equation}}
\nc{\eeq}{\end{equation}}
\nc{\beqa}{\begin{eqnarray}}
\nc{\eeqa}{\end{eqnarray}}
\nc{\bea}{\begin{eqnarray}}
\nc{\eea}{\end{eqnarray}}
\nc{\barray}{\begin{eqnarray}}
\nc{\earray}{\end{eqnarray}}
\nc{\barrayn}{\begin{eqnarray*}}
\nc{\earrayn}{\end{eqnarray*}}
\nc{\ra}{\rightarrow}
\nc{\lsim}{\begin{array}{c}\,\sim\vspace{-21pt}\\< \end{array}}
\nc{\gsim}{\begin{array}{c}\sim\vspace{-21pt}\\> \end{array}}
\nc{\Tr}{{\rm Tr}}
\nc{\slsh}{\slash\hspace*{-0.22cm}}

\def\be{\begin{equation}}
\def\ee{\end{equation}}
\def\bea{\begin{eqnarray}}
\def\eea{\end{eqnarray}}
\def\bit{\begin{itemize}}
\def\eit{\end{itemize}}

\nc{\infinity}{\infty}
\nc{\mc}{\mathcal}
\nc{\M}{\mathcal{M}}
\newcommand{\degree}{^\circ}

\title{
\vspace*{-2.3cm}
\begin{flushright}
\normalsize{
SLAC-PUB-15306\\
  }
\end{flushright}
\vspace{1.5cm}
\Large
\textbf{
Dichromatic Dark Matter
}\vspace*{0.3cm}
}

\author{Yang Bai$^{a,b}$, Meng Su$^{d,e,f}$ and  Yue Zhao$^{b,c}$
\vspace{5mm}
\\
$^{a}$  \normalsize\emph{Department of Physics, University of Wisconsin, Madison, WI 53706, USA}  \vspace{1mm} \\
$^{b}$ \normalsize\emph{SLAC National Accelerator Laboratory, 2575 Sand Hill Road, Menlo Park, CA 94025, USA} \vspace{1mm} \\
$^{c}$ \normalsize\emph{Stanford Institute for Theoretical Physics,
Stanford University, Stanford, CA 94305 USA} \vspace{1mm}\\
$^{d}$ \normalsize\emph{Department of Physics, and Kavli Institute for Astrophysics and Space Research,}   \vspace{1mm} \\
\normalsize\emph{Massachusetts Institute of Technology,
Cambridge, MA 02139, USA} \vspace{1mm} \\
$^{e}$ \normalsize\emph{Institute for Theory and Computation,}\\
\normalsize\emph{Harvard-Smithsonian Center for Astrophysics,
Cambridge, MA 02138, USA} \vspace{1mm}\\
$^{f}$ \normalsize\emph{Einstein Fellow} \vspace{1mm}\\
}

\date{}

\begin{document}
\setcounter{page}{0}
\maketitle

\vspace*{1cm}
\begin{abstract}
Both the robust INTEGRAL 511 keV gamma-ray line and the recent
tentative hint of the 135 GeV gamma-ray line from Fermi-LAT have
similar signal morphologies, and may be produced from the same dark
matter annihilation. Motivated by this observation, we construct a
dark matter model to explain both signals and to accommodate the two
required annihilation cross sections that are different by more than
six orders of magnitude. In our model, to generate the low-energy
positrons for INTEGRAL, dark matter particles annihilate into a
complex scalar that couples to photon via a charge-radius operator.
The complex scalar contains an excited state decaying into the
ground state plus an off-shell photon to generate a pair of positron
and electron. Two charged particles with non-degenerate masses are
necessary for generating this charge-radius operator. One charged
particle is predicted to be long-lived and have a mass around 3.8
TeV to explain the dark matter thermal relic abundance from its late
decay. The other charged particle is predicted to have a mass below
1 TeV given the ratio of the two signal cross sections. The 14 TeV
LHC will concretely test the main parameter space of this lighter
charged particle.
\end{abstract}

\thispagestyle{empty}
\newpage

\setcounter{page}{1}

\baselineskip18pt

\vspace{-3cm}

\section{Introduction}
\label{sec:introduction}
Although dark matter serves as the dominant component of matter in
our universe, its various properties remain unknown. From
astrophysical evidence, there is no doubt that dark matter can
interact with the Standard Model (SM) particles through
gravitational interaction. However, whether there are additional
interactions between dark matter and SM particles is still a mystery
to us. Among several approaches to search for dark matter particles,
measuring the cosmic ray spectrum provides the indirect detection of
dark matter. Observing a high-energy gamma-ray line has long been
believed to be the ``smoking gun" of the dark matter
detection~\cite{Bergstrom:1997fh, Bern:1997ng, Bergstrom:1997fj,
Ullio:1997ke, Perelstein:2006bq, Bertone:2010fn}. Furthermore, the
propagation of energetic photons in our Galaxy is less affected by
the interstellar gas or Galactic magnetic field. The gamma-ray line
signal can even provide the dark matter geometrical profile in our
Galaxy.

The detection of celestial gamma-ray line at $511$ keV from the
inner galaxy, which is believed to be caused by $e^+e^-$
annihilations, was first reported by~\cite{1972ApJ...172L...1J} and
later confirmed
by~\cite{1973ApJ...184..103J,1975ApJ...201..593H,1978ApJ...225L..11L,1979ApJ...228..928B}.
The total flux of the 511 keV line has been estimated to be around
$2\times10^{-3}~{\rm cm^{-2}~s^{-1}}$~\cite{Ascasibar:2005rw}. About
$97\%$ $e^+e^-$ annihilations proceed through the intermediate state
of a positronium atom, and 25$\%$ of these annihilations with
opposite spin of $e^+$ and $e^-$ can produce 511 keV line
emission~\cite{2005MNRAS.357.1377C, 2006A&A...450.1013W}. Although
this gamma-ray line has been known for decades, the identification
of the positron source remains undetermined. Different astrophysical
sources have been suggested during the years, but each of the models
faces various challenge to explain the observations consistently.
The relatively high ratio of the bulge to disk 511 keV emission
towards the inner Galaxy seems against its origin from hypernovae
and gamma ray bursts, while the constraints on the production rate
of high energy positrons also disfavors millisecond pulsars, as well
as proton-proton collisions from e.g. microquasars, low luminosity
X-ray binary jets, and the central supermassive black hole.
Furthermore, pulsars, magnetars, and Galactic cosmic rays are not
favored as major sources to the observed 511 keV from the bulge, and
stringent constraints on these origin of the 511 keV line was
suggested~\cite{Beacom:2005qv,Prantzos:2010wi}.

Besides these astrophysical suggestions, the possibility that dark
matter may create the 511 keV gamma-ray line has been widely
discussed, mainly motivated by the rather spheroidal, symmetric,
bulge-centered morphology.  The lack of higher energy gamma ray
requires the injection energy of positrons to be below $\sim$ 3
MeV~\cite{Beacom:2005qv}.  This motivates studies for both MeV-scale
dark matter
models~\cite{Picciotto:2004rp,Hooper:2004qf,Boehm:2003bt,Pospelov:2008,Hooper:2008,Huh:2008}
and TeV-scale dark matter models with a MeV mass splitting among different dark matter
states~\cite{Pospelov:2007,Finkbeiner:2007kk,Arkani-Hamed:2009,Chen:2009}.
Since TeV-scale dark matter with electroweak interaction strength can
naturally gives correct thermal relic abundance, those models are
more favored. Interestingly,  the morphology of the 511 keV signal
profile has a peaked structure around the Galactic center, and the
sharpness of the peak prefers to have dark matter annihilation
rather than decaying as an explanation~\cite{Ascasibar:2005rw}.
Thus we focus on the heavy dark matter scenario, and try to explain
the $511$ keV INTEGRAL signal via dark matter annihilation.

One popular dark matter model to explain the INTEGRAL signal is the
excited dark matter model with an MeV mass
splitting~\cite{Finkbeiner:2007kk}. This class of models suffer from
the requirement of a large kinematic energy of dark matter to excite
the ground state, hence relying on the Boltzmann tail of dark matter
velocity distribution.  It is under a debate whether the excited
dark matter model can generate enough positrons to explain the large
gamma-ray flux for the INTEGRAL
data~\cite{Chen:2009av,Morris:2011dj}. For the 100 GeV dark matter
mass region that we will consider in this paper, the situation is
worse, because it requires a higher velocity to obtain enough
kinematic energy comparing to a TeV mass dark matter. In our paper,
we will address this problem and propose a new scenario of the
Down-scattering excited Dark Matter (DeDM) to solve the Boltzmann
suppression problem of the vanilla excited dark matter models.

More recently, the hint for another gamma-ray line around 130 GeV
from the Galactic center has been suggested by analyzing the public
data from Fermi Gamma-ray Space Telescope
(Fermi-LAT)~\cite{Weniger:2012tx,Bringmann:2012vr}. The hint becomes
even stronger with the template fitting approach, which takes into
account the spatial distribution of the LAT events towards the inner
Galaxy along with the spectral information~\cite{Su:2012ft}.
Fermi-LAT Collaboration has confirmed the hint of the peak at
$\sim130$ GeV using Pass 7 data. The peak shifts to a higher mass at
$\sim135$ GeV and the significance becomes weaker using the
reprocessed Pass 7 data~\cite{Fermi:2012}. Such high energy
gamma-ray line emission has been considered as a clean signature
from dark matter annihilations. Many dark matter models have been
constructed to explain the 130(135) GeV gamma-ray line feature
(see~\cite{Bringmann:2012ez} and references therein).

The morphology of the INTEGRAL 511 keV and Fermi-LAT 130(135) GeV
line shares similar structures: (1) the signal events concentrate at
the center of the Galaxy with non-disk like distributions; (2) after
smoothing Fermi-LAT signal with INTEGRAL's angular resolution, they
have comparable full widths at half maximum (FWHM) in both the
longitudinal and latitude directions. This motivates us to explain
both signals using same dark matter particle in our universe. The
fittings for both signals prefer annihilation rather than
decaying~\cite{Ascasibar:2005rw, Buchmuller:2012rc}.  Having worked
out the required annihilation cross sections, we find that the
INTEGRAL 511 keV cross section is six to seven orders of magnitude
larger than that of the Fermi-LAT 130 GeV line.  This large
hierarchy of cross sections sets a challenge when constructing a
detailed model. However, the order of magnitude is comparable with
an electromagnetic loop factor of ${\cal O} (\alpha^2/\pi^2)$ if the
INTEGRAL and Fermi-LAT signals are coming from tree-level and
loop-level processes, respectively. This serves as a clue for our
model building.

Our paper is organized as following. In $\S$\ref{sec:morphology}, we
emphasize the similarities of morphologies for both signals and work
out the required cross sections. In $\S$\ref{sec:models}, we propose
our model, Down-scattering excited Dark Matter model, to explain
both signals. In section $\S$\ref{sec:operator}, we first provide a
general operator analysis to illustrate the essence of our model and
calculate the scales of cutoffs of the effective operators. Then we
build up a concrete UV-completion for the operator analysis in
$\S$\ref{sec:renomalizable}.  In $\S$\ref{sec:relic}, we discuss the
thermal history of our model. One charged particle needs to be
long-lived in our UV model, so that we have a semi-natural model to
explain the final dark matter relic abundance.

\section{Experimental Data }
\label{sec:morphology}
In this section, we discuss the INTEGRAL and Fermi-LAT
oberservations in more detail. Since photon is not much affected
during its propagation in the Galaxy, photon coming from dark matter
annihilation can be used to determine the dark matter distribution
in our Galaxy. However, there are subtleties on how to map the
INTEGRAL 511 keV gamma-ray line signal profile to the dark matter
distribution profile. This is because the low-energy positrons that
are generated from dark matter particles can propagate through the
interstellar medium and annihilate with electrons to photons away
from the production site and bias the inferred dark matter
distribution from the 511 keV line morphology. In this paper, we
assume that the positron propagation is negligible comparing to the
spatial resolution of the INTEGRAL, thus the dark matter profile can
be estimated by measuring 511 keV emission morphology. On the other
hand, the Fermi-LAT 130(135) GeV photons could directly be generated
from dark matter particles, and its morphology can therefore tell us
the dark matter profile.

We first compare the morphologies of the INTEGRAL $511$ keV and
Fermi-LAT 130(135) GeV lines.  After smoothing the Fermi-LAT
130(135) GeV line using the angular resolution of INTEGRAL, we find
the spatial distributions are comparable to each other. Furthermore,
assuming that both signals are generated by dark matter
annihilation, we estimate the annihilation cross sections for the
two processes. They will serve as inputs for our model building in
the rest of the paper.

\subsection{Experimental Data} \label{sec:morphologysub1}
\begin{figure}[t!]
\begin{center}
\hspace*{-0.75cm}
\includegraphics[width=0.455\textwidth]{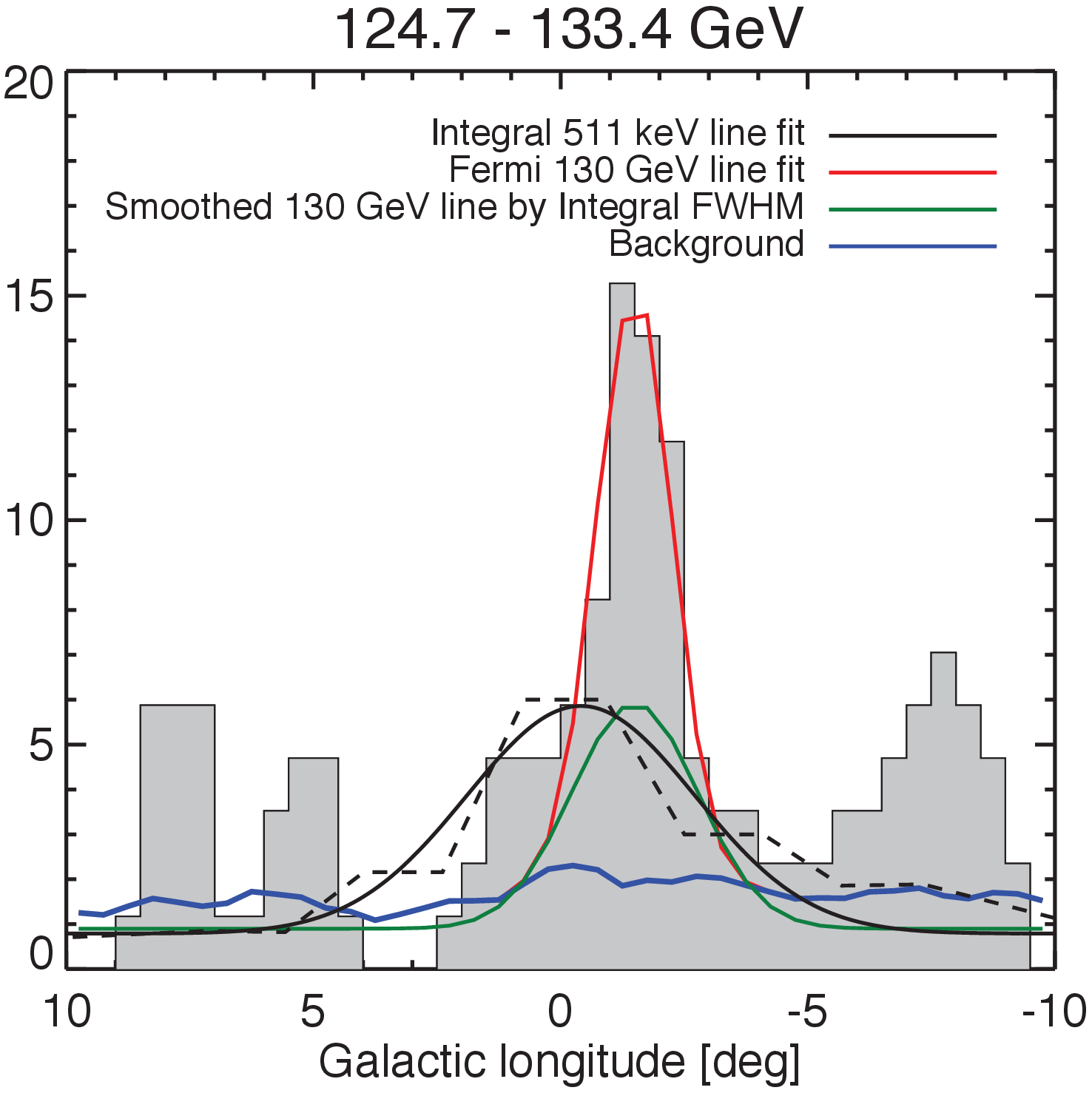} \hspace{1cm}
\includegraphics[width=0.45\textwidth]{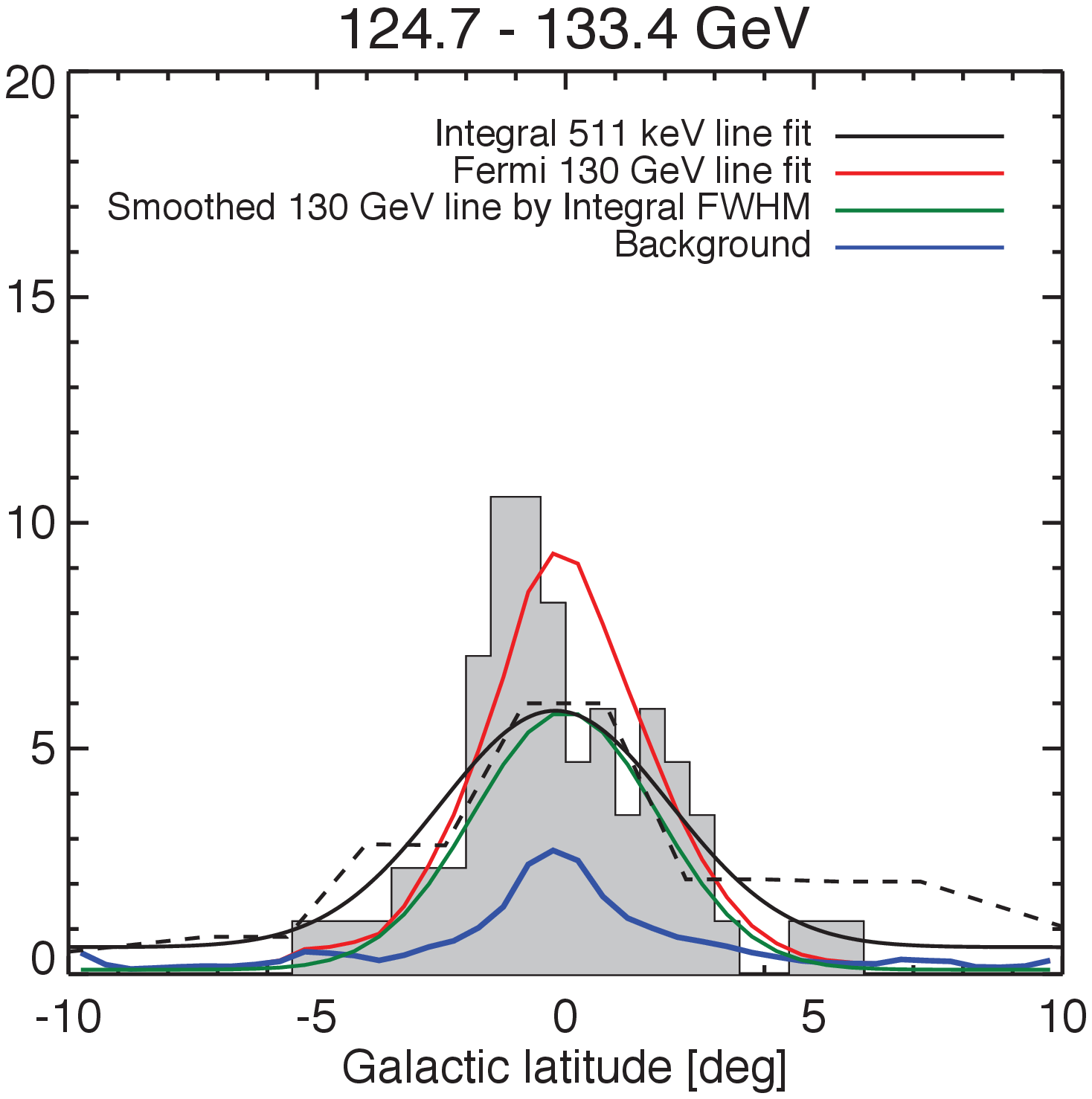}
\caption{Comparison of the INTEGRAL 511 keV line
profile~\cite{Bouchet:2010dj} and Fermi-LAT 130 GeV line
profile~\cite{doubleline}  from the Galacitc center, on longitudinal
(left) and latitudinal (right) projections. The black dashed line
shows the 511 keV line profile measured by INTEGRAL, and the black
solid line shows the fitted Gaussian. The shaded histogram shows the
130 GeV line profile from 3.7 years of Fermi-LAT data. The red solid
line shows the best fitted Gaussian of the 130 GeV line, which is
the same as Fig. 15 in \cite{Su:2012ft}. The green solid line
illustrates the 130 GeV line profile smoothed by SPI 2.6$\degree$
FWHM beam. For the INTEGRAL data, the vertical axis has an arbitrary
scale in this plot.  } \label{fig:compare}
\end{center}
\end{figure}

Thanks to a coded mask located $1.7$ m above the detector plane and
a specific dithering observational strategy, the spectrometer (SPI)
onboard the INTEGRAL observatory can image the sky with a spatial
resolution of $\sim$ 2.6$\degree$ (FWHM).  Based on observations
recorded from February $22nd$, 2003 to January, 2$nd$ 2009, the
study in~\cite{Bouchet:2010dj} has obtained the morphology of the
511 keV line towards the inner Galaxy. In Figure~\ref{fig:compare},
we compare the intensity of the 511 keV gamma-ray line as a function
of Galactic longitude and latitude with the 130 GeV line profile
obtained by fitting 3.7 years Fermi-LAT
observations~\cite{doubleline}. Especially, the dark green line
shows the 130 GeV line profile further smoothed by SPI 2.6$\degree$
FWHM beam.

Interestingly, both longitudinal and latitude distributions of
INTEGRAL are comparable to those of Fermi-LAT after smoothing.
Furthermore, both distributions show the tendency to be off-center
in the negative longitudinal
direction\cite{Su:2012ft,Bouchet:2010dj}. These similarities
motivate the attempts to build one dark matter model to explain both these
two signals.

\subsection{Dark matter annihilation cross sections for INTEGRAL and Fermi-LAT} \label{sec:Xsec}
As discussed in the previous section, both the 511 keV line and the
Fermi-LAT 130(135) GeV line could potentially be explained by dark matter
annihilation. In this section, we estimate the required
annihilation cross sections for both experiments.

The gamma-ray line intensity in a given direction provided by dark matter
annihilation is the line-of-sight integral of the squared dark matter number
density along that given direction
\begin{equation}\label{eq:flux}
\frac{d\Phi_\gamma}{dE_\gamma} = d_{\rm
\chi}\,N_{\rm conv}\,\frac{N_{e^+/\gamma}\,\langle \sigma
v_r\rangle_{e^+/\gamma}}{2} \, \frac{R_\odot \,\rho_\odot^2\,
J}{4\pi\,m_\chi^2}\, \delta (E- E_\gamma)\,,
\end{equation}
with the $J$-factor defined as:
\begin{equation}\label{eq:J}
J = \int db \int d\ell \int \frac{ds}{R_\odot} \, \cos{b}\,
\left[\frac{\rho(r)}{\rho_\odot}\right]^2,
\end{equation}
where $l$ and $b$ are longitude and latitude, and the integral of
$s$ is along the line of sight. Here, $R_\odot \simeq 8.5$~kpc is
the distance from the Sun to the galactic center; $\rho(r)$ is the
dark matter halo profile; $\rho_\odot \simeq 0.4$ GeV cm$^{-3}$ is the
often-used dark matter density in the Solar system~\cite{Jungman:1995df}; the
relation between $r$ and $s$ is $r^2 = s^2 + R_{\odot}^2 - 2 s
R_\odot \cos\ell \cos b$; $N_{\gamma}$ ($N_{e^+}$) is the number of
photons (positrons) generated from each dark matter annihilation hard
process; $m_\chi$ is the dark matter mass; $\langle\sigma v_r\rangle_{e^+}$
and $\langle\sigma v_r\rangle_{\gamma}$ are the annihilation cross
sections. We define $d_{\rm \chi}=1$ for self-conjugated dark matter, e.g. a
real scalar or a Majorana fermion, and $d_{\rm \chi}=\frac{1}{2}$
for a complex scalar or a Dirac fermion. $N_{\rm conv}$ is the
number of monochromatic photons that the final states could convert
to.  For Fermi-LAT, $N_{\rm conv}=1$, since we assume that only
monochromatic photons are produced in the hard process. For
INTEGRAL, observations suggest that about $97\%$ of positrons
annihilate through positronium formation~\cite{Jean:2005af}. Only
1/4 of annihilation takes place in the parapositronium state, which
produces two 511 keV photons. So, we have $N_{\rm conv}\approx0.55$
for INTEGRAL.

We consider both the Einasto and
the Navarro-Frenk-White (NFW) dark matter profile
\beqa
\rho_{\rm _{Ein}}(r) =\rho_\odot\,
e^{-\frac{2}{\alpha}\left[\left(\frac{r}{r_s}\right)^\alpha-\left(\frac{r_\odot}{r_s}\right)^\alpha\right]}\,,
\qquad\qquad
\rho_{\rm _{NFW}}(r) =\rho_\odot\,
\left(\frac{r_{\odot}}{r}\right)^\alpha\,\left[\frac{1+r_\odot/r_s}{1+r/r_s}\right]^{3-\alpha}\,,
\eeqa
with $r_s=20$~kpc and $\alpha=0.17$ for
Einasto~\cite{Navarro:2003ew} and $\alpha=1$ for
NFW~\cite{Navarro:1996gj}. Using the fitted fluxes for the INTEGRAL
signal (the dark matter+disk ones) in Ref.~\cite{Vincent:2012an}, we obtain
the annihilation cross sections as~\footnote{Here, we use different
parameters for dark matter profiles compared to the ones in
Ref.~\cite{Vincent:2012an}. We simply rescale their signal flux by
the ratio of $J$ functions, which could bring an uncertainty of
${\cal O}(1)$.}
\beqa \langle \sigma v_r \rangle_{\gamma,\rm 511,Ein(NFW)}
=\, \frac{1}{d_{\rm\chi}N_{e^+}}\,\times \, 1.5(0.28)\,\times 10^5\times
\left(\frac{m_{\chi} } {100~{\rm GeV} }\right)^2 ~\mbox{pb}\cdot {c}
 \,. \label{eq:xintegralEin511}
\eeqa
For the Fermi-LAT 130(135) GeV gamma line, we use the fitted fluxes
from Ref.~\cite{Weniger:2012tx} for both profiles to calculate the
cross sections,
\beqa \langle \sigma v_r \rangle_{\gamma,\rm 135,Ein(NFW)} =
\, \frac{1}{d_{\rm \chi}N_{\gamma}}\,\times \, 0.42(0.76)\,\times
10^{-1}\times \left(\frac{m_{\chi} } {100~{\rm GeV} }\right)^2
~\mbox{pb}\cdot {c}
 \,.  \label{eq:xfermiEin130}
\eeqa
To quantify the ratio of the required cross sections for two experimental results,
we define $R^{135}_{511} \equiv\langle \sigma v_r \rangle_{\gamma,\rm
135} /\langle \sigma v_r \rangle_{\gamma,\rm 511}$. Taking $N_{e^+}=N_{\gamma}=2$, we have the experimentally measured ratios as
\beqa (R^{135}_{511})_{\rm exp,Ein} \approx 2.9 \times 10^{-7}\,,\qquad\quad
(R^{135}_{511})_{\rm exp,NFW} \approx 2.7 \times 10^{-6}\,, \eeqa
where clearly show a large hierarchy for the two required cross
sections. We want to also stress that the astrophysical
uncertainties are fairly large and a global fit by combining the
INTEGRAL and Fermi-LAT might bring the uncertainty down. The cross
section ratio between these two expertiments is ${\cal O}(10^{-6}
\sim10^{-7})$.  This will be the input for model building in latter
sections. Interestingly, this ratio is comparable to the square of
the electromagnetic loop factor $(\alpha^2/\pi^2)\sim
6\times10^{-6}$, which implies these two experimental results may be
related by a loop with two electromagnetic vertices.  It serves as
a clue for model building.

\section{Down-scattering excited Dark Matter}
\label{sec:models}
There are several interesting features required to construct dark matter
models if both signatures are to be explained by the same dark matter
particle with a mass at the 100 GeV scale.
\bit
\item The required cross section for the INTEGRAL data is amazingly large. For a simplest estimation on the annihilation rate, we get $\sigma v\sim 1/(4\pi \,m^{2}_\chi) \sim 3\times 10^3$~$\mbox{pb}\cdot {c}$ for $m_\chi$ around 100 GeV.  This estimation is three orders of magnitude smaller than the required one. Additional mechanisms are therefore required to increase the annihilation rate. There are several ways to achieve this and we pay special attention on the resonance enhancement~\cite{Ibe:2008ye,Lee:2012,Buckley:2012,Lee:2012wz,Bai:2012qy,Chalons:2012xf}.
\item To explain the INTEGRAL data, primary positron injections from dark matter are required. Since we don't see any excess for other cosmic rays, the underlying dark matter model should be arranged to treat positron/electron differently from other particles. In principle, this can be achieved either from kinematic constraints or symmetry reasons.
\item The ratio of the two cross sections  is $\langle \sigma v_r \rangle_{\rm 135}/\langle \sigma v_r \rangle_{\rm 511} \sim 10^{-7} \ $or$\  10^{-6}$. The dark matter model should also provide a natural explanation for this hierarchy of two cross sections.
\item The model should provide correct amount of dark matter
relic abundance to be consistent with observation. \eit

In the following, we will provide a particle physics model to
incorporate all above four ingredients. Specifically, we will use a
resonance particle in the $s$-channel to increase the dark matter
annihilation cross section required for 511 keV gamma-ray line. The
kinematic constraints from a small mass splitting will be used in this
paper to select positron/electron as the signals from dark matter
annihilation. Instead of introducing a light mediator, e.g. dark
photon, for the dark matter sector connecting to the positron/electron, we
use photon as a more natural mediator to achieve this goal. Noticing
that a neutral scalar field cannot decay into another neutral scalar
field plus one on-shell photon, which is the reason why
$\Upsilon[(n+1)S]\nrightarrow \Upsilon[n S] + \gamma$, a neutral
scalar coupling to photon with the charge-radius operator can
naturally induce $e^{\pm}$ without generating a photon signal in the
meanwhile. The mass difference of the two scalar fields is chosen to
be small such that the kinematic energy of $e^\pm$ is small enough
to be consistent with observation. To explain the ratio of the cross
sections, we will have the cross section for INTEGRAL to be
controlled by coefficients of renormalizable operators, while
loop-generated higher-dimensional operators for Fermi-LAT. We will first
perform an operator analysis and then provide a UV-complete model.

\subsection{Operator analysis}
\label{sec:operator}
We introduce one Dirac fermion $\chi$ and one complex scalar field
$\Phi \equiv (\phi_1 +i \phi_2)/\sqrt{2}$ in the dark matter sector. Both
$\chi$ and $\phi_1$ are stable particles and coexist in our current
Universe. In our study, we will assume that the dark matter component $\chi$
occupies the majority of the dark matter energy, but we will come back to
discuss the relative relic abundances of them later. The
interactions of the dark matter sector to the SM particles are described by
the following set of effective operators
\beqa
-{\cal L} \supset i\,\lambda_\chi\,\overline{\chi} \gamma^5 \chi\,S + \mu\, S\,\Phi^\dagger \Phi  \,+\, \frac{\lambda_S\,\alpha}{4\pi\,M}\,S\,F_{\mu\nu}F^{\mu\nu} \,+\, \frac{\lambda_\Phi\,e}{16\pi^2\,M^2}\,\partial_\mu \Phi \partial_\nu \Phi^\dagger F^{\mu\nu} \,,
\label{eq:operators}
\eeqa
where we implicitly assume that the higher-dimensional operators can
be generated at one-loop level.  The annihilation of $\chi$'s is through
exchanging the real scalar $S$ in the $s$-channel.  For the INTEGRAL data,
a small mass scale at around 1~MeV is required to generate positrons
almost at rest. In our model, we introduce this small mass scale as
the mass splitting of $\phi_1$ and $\phi_2$ from $\Phi = (\phi_1 + \phi_2)/\sqrt{2}$ such that $\delta \equiv
m_{\phi_2} - m_{\phi_1} \ll m_{\phi_1}$ and $\delta \sim 1$~MeV.
Noticing that the parameter $\delta$ explicitly breaks the global
$U(1)_\phi$, so the smallness of $\delta$ is technically natural.
Expanding the last operator in terms of $\phi_1$ and $\phi_2$, we
have
\beqa
\frac{\lambda_\Phi\,e}{16\pi^2\,M^2} \, i\,\partial_\mu \phi_2 \partial_\nu \phi_1 F^{\mu\nu} \,.
\eeqa
Using the equation of motion, one can rewrite the above operator as
$\phi_2\partial{_\nu}\phi_1 \partial_\mu F^{\mu\nu} =
\phi_2\partial{_\nu}\phi_1 j^{\nu}$. This indicates that $\phi_2$
cannot decay to a mass-on-shell photon. For $2\,m_e < \delta <
2\,m_\mu$, we have the leading decay channel of $\phi_2$ as
\beqa
\phi_2 \rightarrow \phi_1 \,+\, \gamma^* \rightarrow \phi_1 \,+\, e^+\,+\, e^- \,.
\label{eq:phi2decay}
\eeqa
Photon, naturally, behaves as a mediator for the dark matter sector to generate
positrons.

\begin{figure}[t!]
\begin{center}
\hspace*{-0.75cm}
\includegraphics[width=0.45\textwidth]{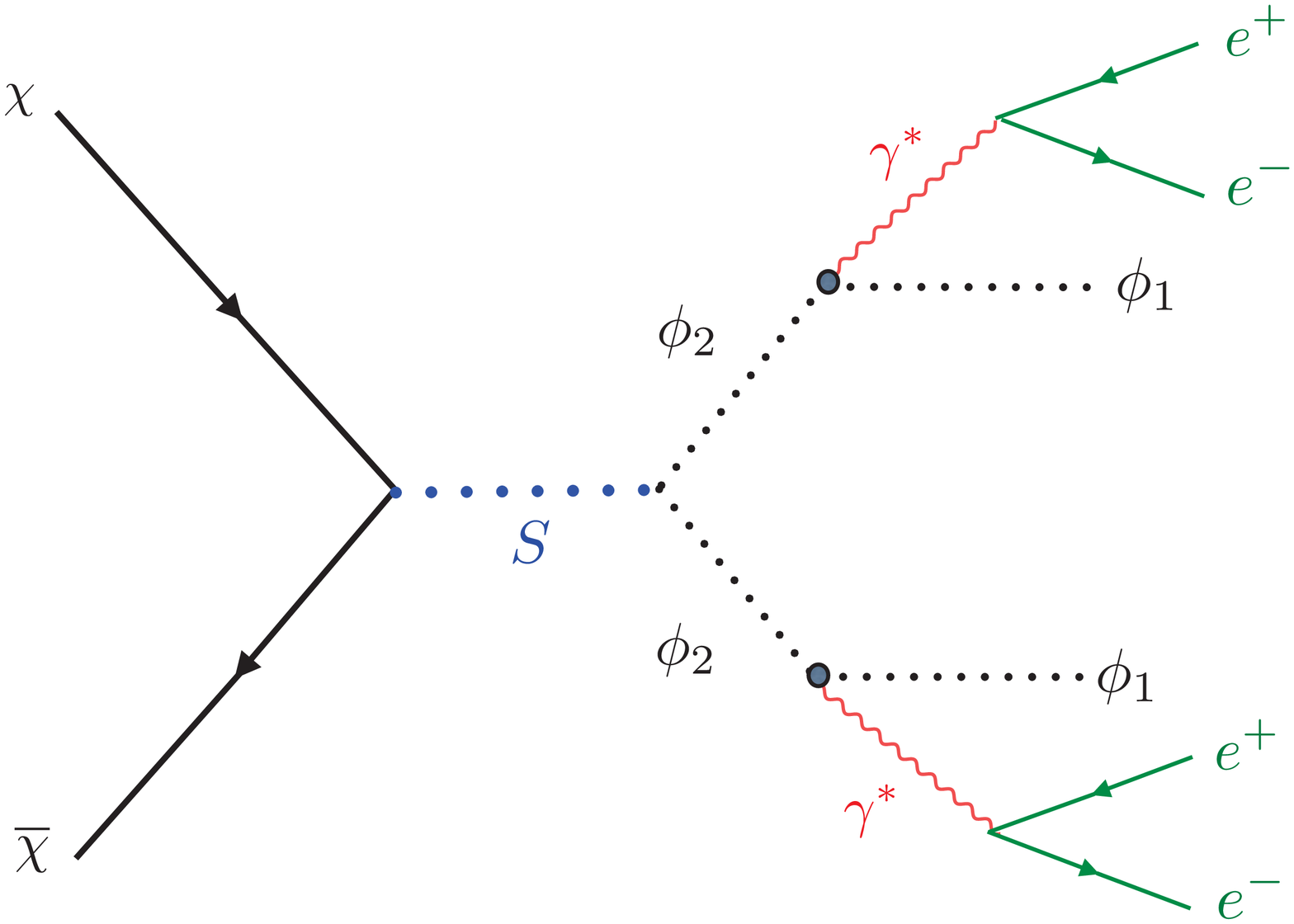} \hspace{1cm}
\includegraphics[width=0.35\textwidth]{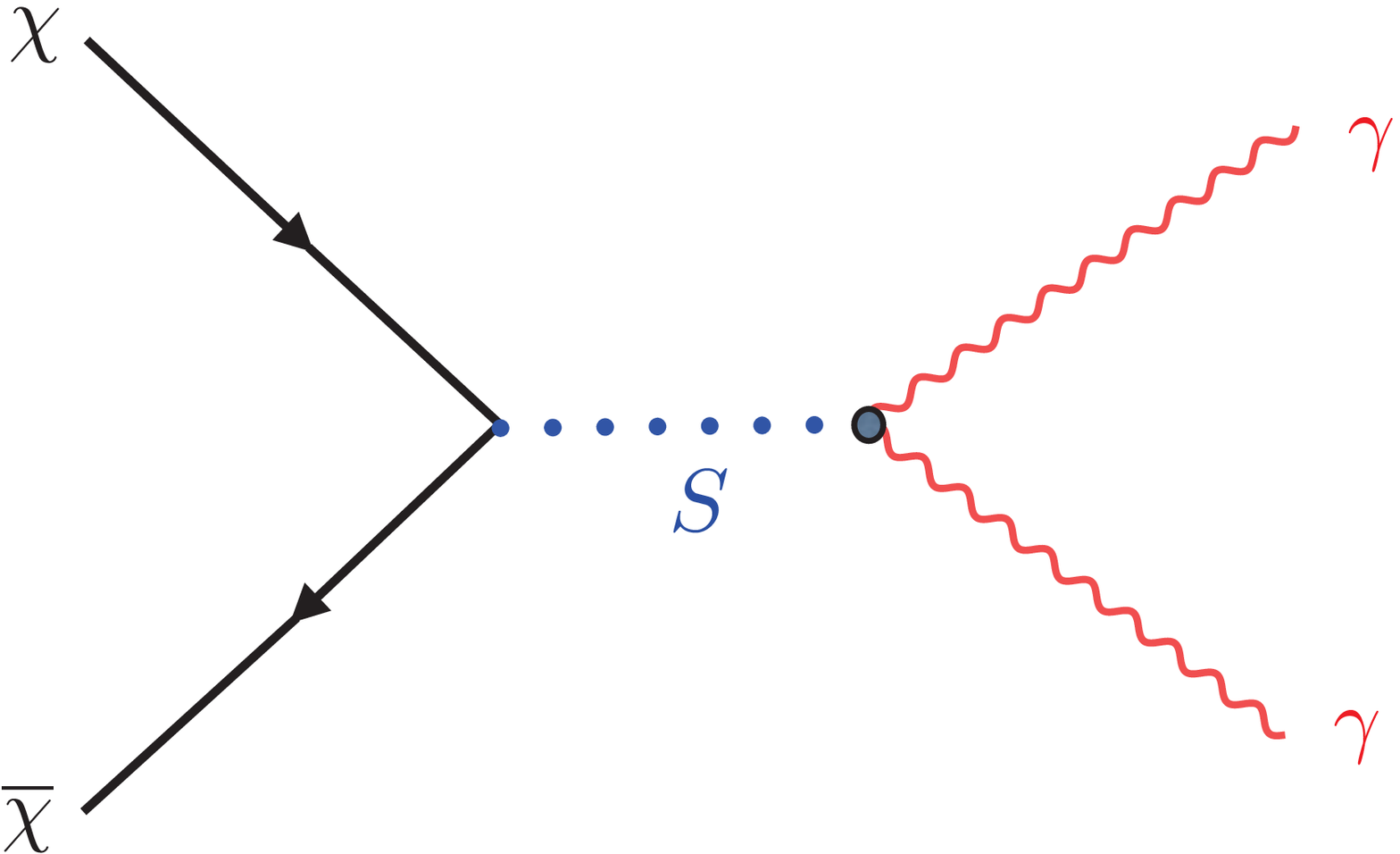}
\caption{The Feynman diagrams for INTEGRAL (left) and Fermi-LAT (right).}
\label{fig:Feyndiagrams}
\end{center}
\end{figure}

The processes to generate positrons for INTEGRAL and photons for
Fermi-LAT are shown in Fig.~\ref{fig:Feyndiagrams}, where the solid
thick points indicate higher-dimensional operators for those
vertices. Although it looks like that the relative cross sections
for those two processes are unrelated to each other, we will show in
a concrete renormalizable model that the overall cross sections
could have a relation in \S \ref{sec:renomalizable}. In order to
generate slowly moving positron from dark matter annihilation, as preferred
from the INTEGRAL data, there are two conditions required: (1) the
mass splitting $\delta$ should be close to $2m_e$; (2) $\phi_2$
cannot have a large boost. The first condition can be satisfied by
choosing $\delta \gtrsim 2m_e$. The second condition can be arranged
by choosing $m_{\phi_2} \lesssim m_\chi$.

We first calculate the annihilation cross section for INTEGRAL. Using the interactions of $\phi_2$ in Eq.~(\ref{eq:operators}), one
gets the annihilation cross section of $\chi \bar{\chi}\rightarrow
\phi_2 \phi_2$ at leading order in $v_r$ as
\beqa
( \sigma v_r)_{\phi_2\phi_2}=\frac{\lambda_\chi^2 \mu^2}{32\pi} \frac{1}{(4m_\chi^2 + m_\chi^2 v_r^2 - m^2_{S})^2+m_{S}^2 \Gamma_{S}^2}\, \sqrt{1 - \frac{m_{\phi_2}^2 }{m_\chi^2}}
\label{eq:xxphi2}
 \,,
\eeqa
We are interested in the parameter space with $2m_\chi > m_S >
2m_{\phi_2}$. The decay width of $S$ is calculated to be $\Gamma_S
\approx 2\,\Gamma^{\phi_2}_S + \Gamma^\gamma_S$ with
\beqa
\Gamma^{\phi_2}_S &=& \frac{\mu^2}{32\pi\,m_S} \sqrt{1\,-\,\frac{4m_{\phi_2}^2}{m_S^2} }   \,, \\
\Gamma^{\gamma}_S &=& \frac{\lambda_S^2\,\alpha^2\, m_S^3}{64\,\pi^3\,M^2} \,.
\label{eq:Swidths}
\eeqa
Here we treat the decay width of $S \rightarrow \phi_1 \phi_1$ to be
approximately the same as $\Gamma^{\phi_2}_S$.

\begin{figure}[t!]
\begin{center}
\hspace*{-0.75cm}
\includegraphics[width=0.45\textwidth]{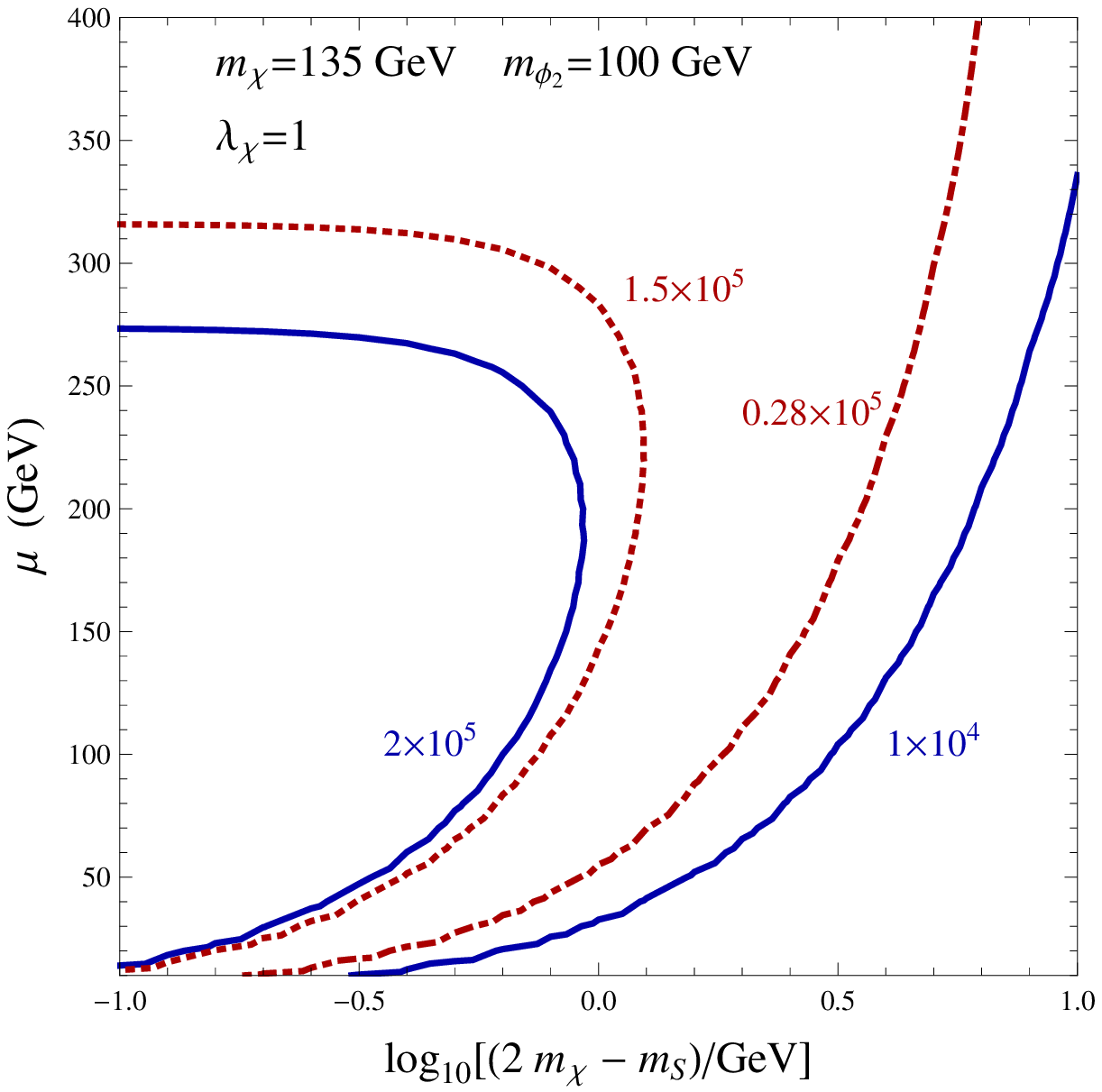} \hspace{1cm}
\includegraphics[width=0.45\textwidth]{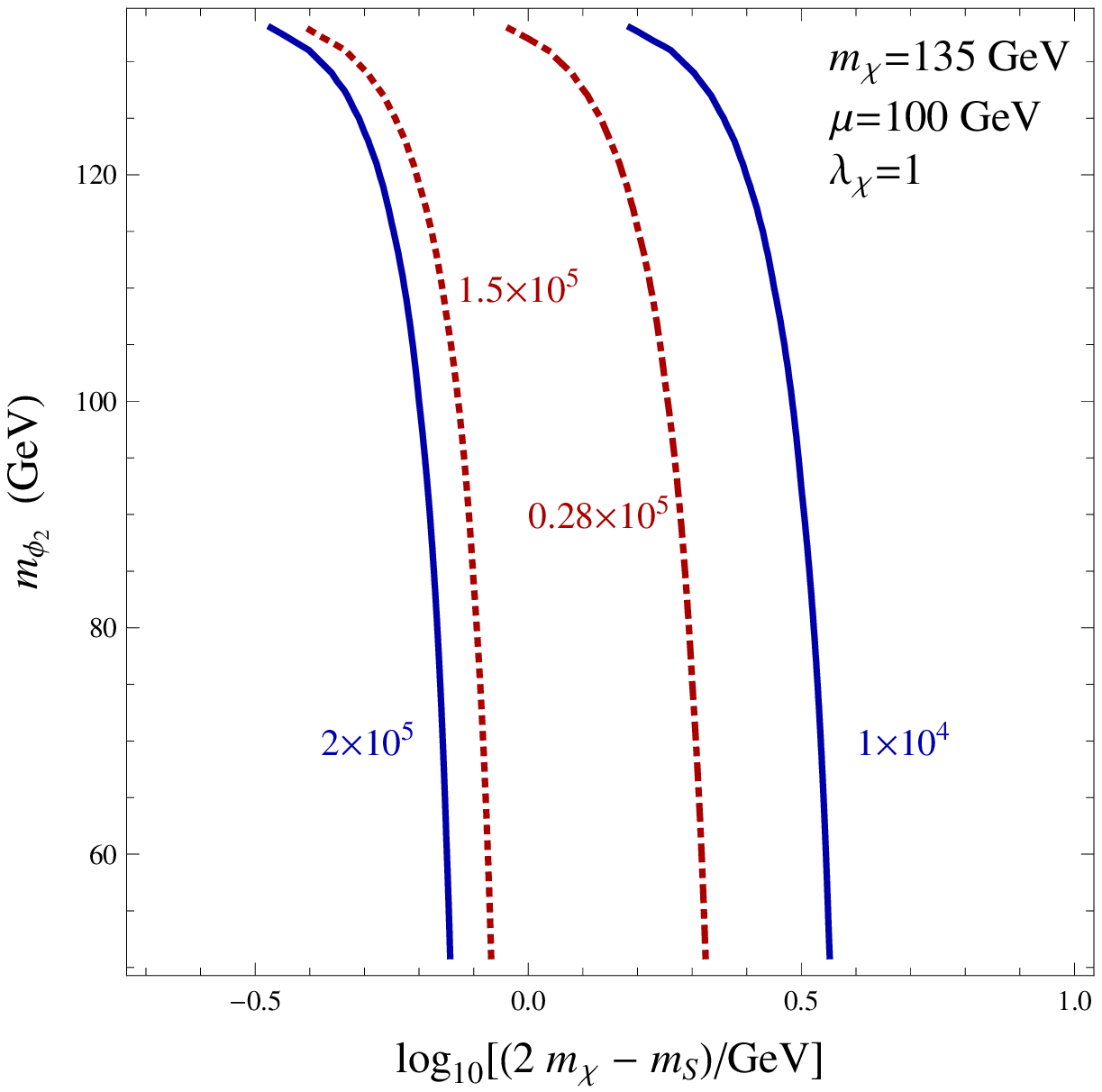}
\caption{Left panel: the contour plots of the annihilation rates in
pb$\cdot$c for $\mu$ and the mass difference $(2m_\chi - m_S)$. The red and dotted line is the required cross section to explain the INTEGRAL data for the Einasto profile, while the red and dotdashed line is for the NFW profile in Eq.~(\ref{eq:xintegralEin511}).  Right panel: the same as the left one but in terms of
$m_{\phi_2}$ and $(2m_\chi - m_S)$.} \label{fig:Integralx}
\end{center}
\end{figure}

For INTEGRAL, we need to calculate the velocity-averaged annihilation rate, which is given by
\beqa
\langle \sigma v_r \rangle_{511} =\left\langle ( \sigma v_r)_{\phi_2\phi_2}(v_r)\right\rangle = \frac{x^{3/2}}{\sqrt{4\pi}} \int v_r^2 dv_r e^{-xv_r^2/4} ( \sigma v_r)_{\phi_2\phi_2}(v_r) \,,
\eeqa
where $x=m_\chi / T = v_0^{-2}$ with $v_0$ determining the variance of
the Gaussian dark matter velocity distribution. In our numerical calculation,
we neglect the upper limit of the integration, which is controlled
by the escaping velocity of dark matter in the galaxy and has only a small
effect on our final results.

In Fig.~\ref{fig:Integralx}, we show the contours of the
annihilation rates of $\langle \sigma v_r \rangle_{511}$ in terms of
$\mu$ and $(2m_\chi -m_S)$ in the left panel, also $m_{\phi_2}$ and
$(2m_\chi -m_S)$ in the right panel. To obtain a large annihilation
rate around $10^5$~pb to explain the INTEGRAL data, the resonance mass has to be very close to twice of the dark matter mass. The mass splitting should be a few GeV for the
parameter $\mu\sim 100$~GeV. In Fig.~\ref{fig:Integralx}, we only
presented the results for $m_S \lesssim 2 m_\chi$. The case with
$m_S \gtrsim 2 m_\chi$ has an additional contribution to the total
width of $S$ from $S\rightarrow \chi \bar{\chi}$, and has similar
results. From the right panel of Fig.~\ref{fig:Integralx}, we can
see that the annihilation rate is insensitive to $m_{\phi_2}$ except
for the region with $m_{\phi_2}\sim m_\chi$. One might think that
$\phi_2$ can be as light as possible. However, a light $\phi_2$
generated from dark matter annihilation can have a large Lorentz boost. As a
consequence, $e^+$ from $\phi_2$ decays is also boosted and too
energetic to explain the INTEGRAL data~\cite{Ascasibar:2005rw}.
Therefore, we restrict the parameter space in our later study to
have $m_{\phi_2}$ at least above 50 GeV.

\begin{figure}[t!]
\begin{center}
\hspace*{-0.75cm}
\includegraphics[width=0.55\textwidth]{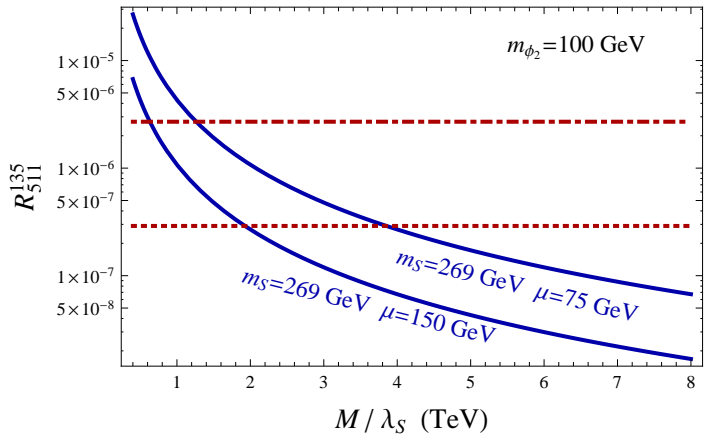}
\caption{The ratio of annihilation rates for the Fermi-LAT and
INTEGRAL signals as a function of the cutoff of the
higher-dimensional operators. Here, we choose $m_{\phi_2}=100$~GeV. The dotted and dot-dashed lines
(red) indicate the approximate value from experimental
measurements for two different dark matter profiles.} \label{fig:Pratio}
\end{center}
\end{figure}

For the Fermi-LAT signal, instead of obtaining the absolute
annihilation rate, we calculate the ratio of the required signal
strengths for Fermi-LAT and INTEGRAL. The ratio is equal to
the branching ratio of the two decay channels of $S$ in
Eq.~(\ref{eq:Swidths}), assuming that the additional contribution
from the process $\phi_1 \phi_1 \rightarrow \gamma \gamma$ is small.
By taking the ratio, the dependence on the resonance propagator is
cancelled and we have
\beqa
R^{135}_{511} = \frac{\lambda_S^2\,\alpha^2\,m_S^4}{2\,\pi^2\,M^2\,\mu^2}\, \left(1\,-\,\frac{4m_{\phi_2}^2}{m_S^2} \right)^{-1/2} \,.
\eeqa
We show this ratio of the annihilation rates in
Fig.~\ref{fig:Pratio} by fixing $m_{\phi_2}=100$~GeV. From
Fig.~\ref{fig:Pratio}, we can find that the cutoff of the operators
in Eq.~(\ref{eq:operators}) should be several TeV.

From the effective operator analysis in this section, we have seen
that it is possible to explain the required annihilation rates for
both INTEGRAL and Fermi-LAT. Our model is economical in a sense that
only a few operators and a small number of degrees of freedom are
required to explain the data. On the other hand, we should also
admit that the resonance requirement of $(2m_\chi - m_S) \ll m_S$ is
a tuning point of the parameter space of the current model.
Additional ingredients are therefore required to explain this
delicate mass relation.  We leave this direction of exploration to a
future study. Here we emphasize that the ratio of INTEGRAL and
Fermi-LAT signals are independent on the way we enhance the
annihilation cross sections.  Thus one can attach the rest of the
model to any other ways of enhancement, e.g. a light mediator in the
$t$-channel plus the Sommerfeld
enhancement~\cite{ArkaniHamed:2008qn}. In the next section, we
construct a renormalizable model to UV complete the Lagrangian in
Eq.~(\ref{eq:operators}) and explain the common origin of the last
two operators in Eq.~(\ref{eq:operators}).

\subsection{Renormalizable model}
\label{sec:renomalizable}
One way to UV complete the effective Lagrangian in the previous
section is to introduce electromagnetic charged states to connect
the dark matter sector to photon. In order to have the state $\phi_1$ stable,
at least two charged particles are required to preserve the discrete
symmetry associated with $\phi_1$. As one example, we introduce two
charged complex scalar fields, $X_1$ and $X_2$. One could also study
fermionic charged states in the same procedure. Under $U(1)_Y$ or
$U(1)_{em}$ after electroweak symmetry breaking, $X_1$ and $X_2$
have charge one. The global symmetries that we introduce contain a
${\cal Z}_2$ symmetry responsible for the stability of the dark matter
particles and a $U(1)_\phi$ protecting the mass degeneracy of
$\phi_1$ and $\phi_2$. We show the field content and symmetries in
Table~\ref{tab:UVmodel}.
\begin{table}[t!]\small
\vspace*{4mm}
\renewcommand{\arraystretch}{1.5}
\centerline{
\begin{tabular}{c|cccc}
\hline \hline
&$\mbox{spin}$ &$U(1)_Y$&$U(1)_{\phi}$ &$ \mathbb{Z}_{2}$\\ \hline
$\chi $&$\frac{1}{2}$&0&0&$-$\\
$S$&0&0&0&$+$\\
$\Phi=\frac{1}{\sqrt{2}}(\phi_1+i \phi_2)$&0&0&1&$-$\\
$X_1 $&0&1&1&$-$\\
$X_2 $&0&1&0&$+$  \\ \hline \hline
\end{tabular}
} \caption{Matter content and corresponding charge assignments. The
global symmetry $U(1)_\phi$ is only an approximate one. The small
mass splitting of $\phi_1$ and $\phi_2$ breaks it.
\label{tab:UVmodel}}
\end{table}

Based on the symmetries in Table~\ref{tab:UVmodel}, we have the
following subset of operators allowed by the symmetries,
\beqa
-{\cal L} &\supset& i\,\lambda_\chi\,\overline{\chi} \gamma^5 \chi\,S + \mu\, S\,\Phi^\dagger \Phi \,+\,
 \mu_1 (\Phi X_1 X_2^\dagger + \Phi^\dagger X_1^\dagger X_2)
  \,+\, \mu_2\,S \,X_1 X_1^\dagger \,+\, \mu_3\,S\, X_2 X_2^\dagger \, +\lambda_1\, \Phi \Phi^\dagger H H^\dagger  \nonumber \\
  &&      \,+\, \frac{1}{2}m_S^2\,S^2\,+\, m_\phi^2 \Phi \Phi^\dagger   \,+\, \frac{1}{2}m_\phi\,\delta\,( \Phi^2 + \Phi^{\dagger\,2}) \,+\, m_{X_1}^2 X_1 X_1^\dagger \,+\, m_{X_2}^2 X_2 X_2^\dagger \nonumber \\
  &&  \,+\, \lambda^i_e\,X_1\,\overline{\chi}\,e_R^i + h.c.
 \,.
\label{eq:operatorsUV}
\eeqa
Here we only list the operators which are relevant to the processes
we concern in this paper.  Especially the operator $SHH^\dagger$ is
neglected, which is assumed to have a small coefficient. The last
operator generically introduces lepton flavor violation processes,
so the couplings $\lambda^i_e$ ($i$ is the flavor index) should be
small.

We first note that the vertices $(\Phi X_1 X_2^\dagger +
\Phi^\dagger X_1^\dagger X_2)$ could generate the charge radius
operator for the $\Phi$ field as shown in the last operator in
Eq.~(\ref{eq:operators}). After a calculation of the triangle
diagram with $X_1$ and $X_2$ propagating in the loop, we get the
Feynman rule of the following operator $\partial_\mu \Phi
\partial_\nu \Phi^\dagger F^{\mu\nu}=i\,\partial_\mu \phi_2
\partial_\nu \phi_1 F^{\mu\nu}$
\beqa
\frac{e\,\mu_1^2}{32\,\pi^2} \int^1_0 dx \int^{1-x}_0 dy \left[ \frac{(1-2x)\,k_1^\alpha + (1-2y)\,k_2^\alpha}{ (1-x-y)m_{X_2}^2 + (x+y)\,m_{X_1}^2 + (x\,k_1 + y\,k_2)^2 - x\,k_1^2 -y\,k_2^2} - (m_{X_1} \leftrightarrow m_{X_2}) \right] \,,
\eeqa
where $k_1$ and $k_2$ are momenta of $\phi_1$ and $\phi_2$ with
opposite directions towards the vertex. In the limit $m^2_{X_{1,
2}}\gg k_1^2, k_2^2$, we can match to the coefficient of the
effective operator in Eq.~(\ref{eq:operators}) as
\beqa
\frac{\lambda_\Phi}{M^2} &=& \,\frac{\mu_1^2\left[3(m_{X_1}^4 - m_{X_2}^4)
- 2 ( m_{X_1}^4 + 4 m_{X_1}^2 m_{X_2}^2 + m_{X_2}^4 ) \log{(m_{X_1}/m_{X_2})}  \right]
}{6(m_{X_1}^2 - m_{X_2}^2)^4} \,.
\eeqa
We notice that the above formula vanishes when $m_{X_1} = m_{X_2}$.
This can be understood by the enhanced discrete symmetry, $\Phi
\rightarrow \Phi^\dagger, X_1 \leftrightarrow X_2$, in the
Lagrangian when $X_1$ and $X_2$ have degenerate
masses.\footnote{Operator $\lambda_e^i X_1 \bar{\chi} e^i_R$ does
not preserve this symmetry, but this operator could have a very small coefficient and is irrelevant to this
calculation.} The charge-radius operator violates this discrete
symmetry, thus cannot be generated when $m_{X_1} = m_{X_2}$. Another
more intuitive explanation is to think $\Phi$ as a composite
particle of $X_1^+$ and $X_2^-$. If the mass of $X_2^-$ is much
heavier than $X_1^+$, one can treat $X_1^+$ as a particle rotating
around $X_2^-$ and have a nonzero charge radius. However, for the
mass degenerate case, $X_1^+$ and $X_2^-$ should be treated with
equal foot and rotate around the center with the same radius. As a
result, for each orbit the net charge is zero and the charge radius
is zero.

Similarly, we can integrate out $X_1$ and $X_2$ to generate the
effective operator coupling $S$ to two photons. To match the
coefficient in Eq.~(\ref{eq:operators}), we have
\beqa
\frac{\lambda_S}{M}&=& \frac{1}{12}\left( \frac{\mu_2}{m_{X_1}^2} + \frac{\mu_3}{m_{X_2}^2} \right) \,.
\label{eq:X1X2mass}
\eeqa
In the limit $m^2_{X_2}/\mu_3 \ll m^2_{X_1}/\mu_2$, we have
\beqa
m_{X_2} \approx 410~\mbox{GeV} \times \left( \frac{M/\lambda_S}{2~\mbox{TeV}} \right)^{1/2}\,\left(\frac{\mu_3}{1~\mbox{TeV}}\right)^{1/2} \,.
\eeqa
Using the values of $M/\lambda_S$ in Fig.~\ref{fig:Pratio}, we
anticipate at least the charged particle $X_2$ to have a mass below
1 TeV. This charged particle $X_2$ can decay into one lepton plus
one neutrino, for example via the higher dimensional operator $X_2
\tilde{H} \bar{L}\,e_R$.

\subsection{Dark Matter Relic Abundance}
\label{sec:relic}
In our DeDM model, we have two stable particles in our spectrum:
$\chi$ and $\phi_1$. In our previous analysis, we have assumed that
the majority of dark matter in our universe is mainly composed of $\chi$. To
justify our assumption, it is important to study the thermal history
of $\chi$ and $\phi_1$. In this section, we demonstrate that our
setup contains enough ingredients to induce a right relic abundance
for $\chi$, thus it could be the dominant part of the dark matter in our current Universe.

The thermal relic abundance of $\phi_1$ is controlled by the
parameter $\lambda_1$ in Eq.~(\ref{eq:operatorsUV}), which is
similar to the ``Higgs portal" dark matter
models~\cite{Shrock:1982kd,Barger:2007im}. For $m_{\phi_1} <
m_h$, the main annihilation cross section is~\cite{Burgess:2000yq}
\beqa
\sigma v_r(\phi_1) &=&\frac{2\,\lambda_1^2\,v_{\rm EW}^2}{(4\,m_{\phi_1}^2 - m_h^2)^2 + m_h^2\,\Gamma_h^2(m_h)  }\,\frac{\Gamma_h(2m_{\phi_1})}{2\,m_{\phi_1}}  \,,
\eeqa
where $v_{\rm EW}=246$~GeV is the electroweak vacuum expectation
value. The function $\Gamma_h(m)$ is the width of a Higgs boson in
the SM with a mass at $m$. For $\lambda_1=1$, $m_{\phi_1}=100$~GeV
and $m_h = 125$~GeV, we have $\sigma v_r(\phi_1)\approx 581$~pb and
$\Omega_{\phi_1} h^2 \approx 1.4\times 10^{-3}\times \Omega_{\rm DM}
h^2$.  Thus the relic abundance of $\phi_1$ can be naturally small.

To satisfy the dark matter relic abundance, a non-trivial thermal history of
$\chi$ is needed. This is because a large annihilation cross section
in Eq.~(\ref{eq:xxphi2}) is needed to explain the INTEGRAL data. The
thermal relic abundance of $\chi$ is very small compared to the
required dark matter energy density. Noticing that the last operator in
Eq.~(\ref{eq:operatorsUV}) can introduce the decay channel, $X_1^+
\rightarrow \chi \,e^+$, the late decay of thermally abundant
$X^+_1$ particles can generate enough $\chi$, and therefore explain why
$\chi$ could be the majority of dark matter.

We first calculate the thermal relic abundance of the charged
particle $X^{\pm}_1$ before it decays into $\chi$ and a
positron/electron. There are two classes of annihilation channels
for $X^{\pm}_1$. The first class has a photon or $Z$ boson
exchanging in the $s$-channel with final states as a pair of the SM
fermions,  $W^+\,W^-$ gauge bosons, and $h\,Z$.  The second class
includes the $t$-channel diagrams, interfering with seagull diagrams.
Assuming that the mass $X_1^{\pm}$ is far above the SM particle
masses and neglecting the SM particle masses, we have
\beqa
\sigma v_r (X_1^+ X_1^- \rightarrow \gamma \,\gamma) &=& \frac{e^4 }{8\,\pi\,m_{X_1}^2}\,,\qquad
\sigma v_r (X_1^+ X_1^- \rightarrow Z \,Z) = \frac{e^4\,s_W^4}{8\,\pi\,c_W^4\,m_{X_1}^2}\,,\; \\
\sigma v_r (X_1^+ X_1^- \rightarrow \gamma \,Z) &=& \frac{3\,e^4\,s_W^2}{4\,\pi\,c_W^2\,m_{X_1}^2}\,,\qquad
\sigma v_r (X_1^+ X_1^- \rightarrow h\,Z) = \frac{e^4 }{1536\,\pi\,c_W^4\,m_{X_1}^2} \,v_r^2\,,
\eeqa
\beqa
\sigma v_r (X_1^+ X_1^- \rightarrow W^+ W^-) &=& \frac{e^4 }{1536\,\pi\,c_W^4\,m_{X_1}^2} \,v_r^2\,,\; \\
\sigma v_r (X_1^+ X_1^- \rightarrow \bar{f} f) &=& \frac{e^4\,n^f_c\,\left[ c_W^2 - 2 c_W s_W\,q^f\, g_V^f  + s_W^2 ( g_A^{f\,2} \,+\, g_V^{f\,2} )\right]}{96\,\pi\,c_W^2\,m_{X_1}^2} \,v_r^2\,,
\eeqa
Here, $q^f$ is the electric charge of the SM fermion;
$g_A^f\,(g_V^f)$ is the axi-vector (vector) couplings of the $Z$ to
the SM fermion up to the electric coupling $e$; $n^f_c =3$ for
quarks and 1 for leptons. To derive the above formulas, we have only
included the leading terms in $v_r$ for each equation. If the
charged particle could occupy the total energy of dark matter in our
universe, $\Omega_{\rm DM}h^2=0.11$, the required mass is calculated
to be $m_{X_1}\approx 750$~GeV. To derive this mass, we have found
that the $p$-wave suppressed annihilation cross section or the terms
at ${\cal O}(v_r^2)$ is subdominant compared to the total
annihilation cross section.

For a lifetime of $X_1^\pm$ not too long in the cosmological time
scale, we should anticipate that $X_1^+$ has already decayed into
its daughter particle and the final dark matter in our current
universe is composed of $\chi$. On the other hand, the lifetime of
$X_1^\pm$ can not be too short. Otherwise, the produced $\chi$
particles from $X_1^\pm$ decays in the early universe can easily
annihilate away and do not provide enough dark matter energy
density. To calculate the thermal history of the $\chi$ field, one
needs to solve for the following coupled Boltzmann equations between
$\chi$ and $X_1^\pm$
\beqa
 \frac{d n_{X_1}}{d t}\,+\,3\,H\,n_{X_1}&=& -\langle\sigma
v\rangle_{X_1}\left(n_{X_1}^2-n_{X_1}^{{\rm eq}\,2}\right)  \,-\,  n_{X_1} \,\Gamma_{X_1}\,, \\
\frac{d n_{\chi}}{d t}\,+\,3\,H\,n_{\chi} &=& \,-\,\langle\sigma v\rangle_{\chi}\left(n_{\chi}^2-n_{\chi}^{{\rm eq}\,2}\right) \,+\, n_{X_1}\,\Gamma_{X_1}\,.
\label{eq:thermal1}
\eeqa
Here, in the radiation dominated era, $H=(8\pi \rho/3M_{\rm Pl})^{1/2}$, $t=1/(2H)$, $\rho(T)=g_*\,\pi^2\,T^4/30$, and $n^{{\rm eq}}_i(T) = g_i (m_i T/2\pi)^{3/2} e^{-m_i/T}$, where $g_*=86.25$ is the number of degrees of relativistic freedom and $g_\chi=4$ and $g_{X_1}=2$. It is convenient to rescale the number density by the entropy and to define the quantity $Y_i \equiv n_i /s$ with $s=2\pi^2 g_* T^3/45$. The coupled equations become
\beqa
\frac{d Y_{X_1}}{d x}&=&-\frac{s\,\langle\sigma
v\rangle_{X_1}}{H x}\,\left(Y_{X_1}^2-Y_{X_1}^{{\rm eq}\, 2}\right)  \,-\,\frac{\Gamma_{X_1} Y_{X_1}}{H
x}\,,
\label{eq:therma21}
 \\
\frac{d Y_{\chi}}{d x}&=&\,-\,\frac{s\,\langle\sigma v\rangle_{\chi}}{H
x}\left(Y_{\chi}^2-Y_{\chi}^{{\rm eq}\, 2}\right) \,+\, \frac{\Gamma_{X_1} Y_{X_1}}{H
x}\,,
\label{eq:thermal2}
\eeqa
where $x\equiv m_\chi/T$ and $dx/dt=Hx$. The final $\chi$ relic
abundance is given by $\Omega_\chi=\rho_\chi/\rho_c$, where
$\rho_c=3H_0^2 M^2_{\rm Pl}/8\pi = 1.0539\times 10^{-5}
h^2~\mbox{GeV\,cm}^{-3}$ is the critical density corresponding to a
flat universe and $\rho_\chi=m_\chi s_0 Y_\chi({\infty})$ with
$s_0=2889.2\,\mbox{cm}^{-3}$ being the entropy today.

\begin{figure}[t!]
\begin{center}
\hspace*{-0.75cm}
\includegraphics[width=0.55\textwidth]{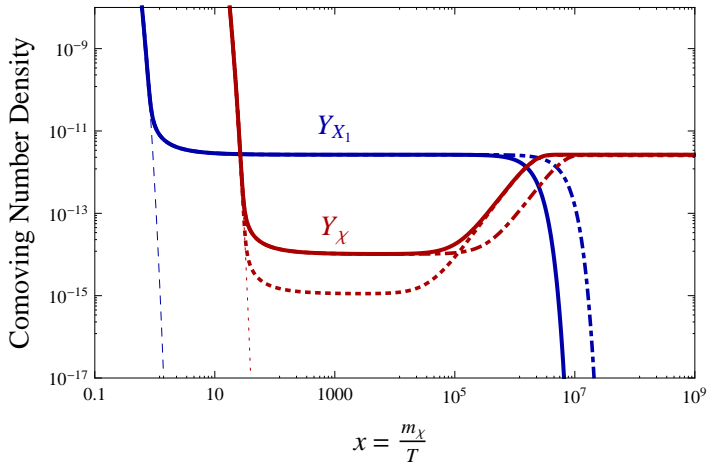}
\caption{The comoving number density as a function of the temperature. Here, we choose $m_\chi=135$~GeV and $m_{X_1}=3.8$~TeV. The annihilation cross section of $X_1$, approximately 0.03~pb, is determined by its interactions with electroweak gauge bosons. The solid lines are for $\langle \sigma v_r \rangle_\chi = 10$~pb and $\tau_{X_1}=10$~s; the dotdashed lines are for $\langle \sigma v_r \rangle_\chi = 10$~pb and $\tau_{X_1}=100$~s; the dotted and red line is for $\langle \sigma v_r \rangle_\chi = 100$~pb and $\tau_{X_1}=10$~s. The relic abundance of $\chi$ satisfies the observed dark matter energy density, $\Omega_\chi h^2 = 0.11$.}
\label{fig:thermal}
\end{center}
\end{figure}

At the temperature region with ${\cal O}(20) < x < 1000$, the decaying terms in Eq.~(\ref{eq:therma21}) and (\ref{eq:thermal2}) are not important. The number densities of $X_1$ and $\chi$ reach their separate freeze-out values. Since the cross section of $\chi$ is much larger than $X_1$, the freeze-out number density for $\chi$ is much below the one of $X_1$. At a later time, only the last terms in Eq.~(\ref{eq:therma21}) and (\ref{eq:thermal2}) become important. One can easily show that the quantity $Y_{X_1} + Y_{\chi}$ is a conserved number. As a result, the final number density of $\chi$ should just match to the number density of $X_1$ at ${\cal O}(20)$. So, approximately we have the relic abundance of $\chi$ as
\beqa
\Omega_\chi h^2 \approx 0.11\times \frac{m_\chi}{135~\mbox{GeV}}\,\times \frac{m_{X_1}}{3.8~\mbox{TeV}}  \,,
\eeqa
So, the charged particle $X_1$ is predicted to be 3.8 TeV and the other charged particle $X_2$ should be below around 1 TeV to explain the ratio of INTEGRAL and Fermi-LAT cross sections in Eq.~(\ref{eq:X1X2mass}).

We solve the coupled equations in Eq.~(\ref{eq:thermal2})
numerically and show both comoving number densities of $\chi$ and
$X_1$ in Fig.~\ref{fig:thermal}. We find that if the mass of $X_1$
is 3.8 TeV, it generates the relic abundance for $\chi$ which
satisfies the total dark matter energy density, $\Omega_\chi
h^2=0.11$. In the blue solid and the red solid lines, for $\langle
\sigma v_r \rangle_\chi = 10$~pb~\footnote{The annihilation cross
section of $\chi$ is not necessarily related to its annihilation
cross section at the current time. This is because its main
production here is from the heavy particle $X_1$ decay, and it has a
relativistic velocity and hence a smaller cross section.} and
$\tau_{X_1}=10$~s we show the evolutions of the $X_1$ and $\chi$
comoving number densities as a function of temperature. As can be
seen from Fig.~\ref{fig:thermal} and at $x\approx 20$, both $X_1$
and $\chi$ have reached ordinary relic abundances according to their
respective annihilation cross sections. At $x\approx 10^5-10^6$,
$X_1$ starts to decay and its number density drops rapidly.
Meanwhile, the stable $\chi$ particle number density increases and
reaches a plateau at around $10^7$. The final number density of the
$\chi$ field is found to be independent on the lifetime
$\tau_{X_1}$, as long as the decay happens late enough so that the
annihilation of $\chi$ is not important any more.  The actual time
for $\chi$ to reach its eventual number density is proportional to
$\sqrt{\tau_{X_1}}$.

To satisfy the dark matter relic abundance, the lifetime of the charged particle $X_1$ can be $\leq 100\,s$ . For such a late decayed particle, we need to worry about its modification on the Big Bang nucleosynthesis (BBN) history. Since the main decaying product of $X_1$ is into leptons plus the stable $\chi$, the BBN constrains are fairly weak. From Ref.~\cite{Jedamzik:2006xz}, the $^6\mbox{Li}/^7\mbox{Li}$ ratio constrains the lifetime of $X_1$ to be $\tau_{X_1} < 10^5\,s$ for $\Omega_{X_1}h^2\approx 0.1$ if it would have not decayed. As pointed in Ref.~\cite{Pospelov:2006sc, Kohri:2006cn}, the long-lived charged particle, with a lifetime $\tau_{X_1}>10^3\,s$, can form a bound state with nuclei and enhance the $^6$Li production. The parameter space in our model can indeed satisfy the BBN constraints.

\section{Discussion and conclusions}
\label{sec:conclusion}
The charged particle $X_1$ in our model behaves as a heavy stable
charged particle (HSCP) at colliders. The current searches from CMS
at $\sqrt{s}=7$~TeV and 5.0 fb$^{-1}$ have set a lower limit on the
mass of $X_1$ to be 223 GeV at 95\% C.L.~\cite{Chatrchyan:2012sp}.
For the predicted mass of $X_1$ around 3.8 TeV, the existing studies
have shown that the 14 TeV LHC with 100 fb$^{-1}$ can reach the HSCP
up to a mass around 1 TeV. So, unlikely the stable charged particle
can be discovered at the 14 TeV LHC. However, for the other charged
particle $X_2$ its mass should be below 1 TeV and could be a
long-lived particle or decay into SM particles, for instance $X_2^+
\rightarrow e^+ \nu_e$. The parameter space of the $X_2$ particle
will be well explored at the  LHC 14 TeV running.

One feature of our model is directly using photon as a mediator to
link the dark matter sector to positron/electron. Unfortunately,
other than searching for the charged particles responsible for the
charge radius operator,  in the near future there is no additional
observable dark matter direct or indirect signatures for the $\chi$
field, which has interactions with SM particles suppressed by the
TeV scale cutoff of the effective operators.  The minor component of
dark matter, $\phi_1$, may have detectable effects.  However, that
highly relies on the parameters one chooses, thus we do not pursue
that in detail here.  Another ingredient that we utilize is the
$s$-channel resonance particle to increase the annihilation cross
section. We want to stress that this option is not a unique one and
is introduced just for convenience. One can also introduce a light
mediator in the $t$-channel plus the Sommerfeld enhancement to
achieve the same goal~\cite{ArkaniHamed:2008qn}.

In summary, we have constructed a realistic model to have the same
dark matter particle responsible for both the INTEGRAL 511 keV and
Fermi-LAT 135 GeV lines. Through an $s$-channel resonance, the dark
matter particles annihilate into a complex scalar, which couples to
photon via a charge-radius operator. For a few MeV mass splitting
between the real and imaginary parts of the complex scalar, two
pairs of electron and positron are the main visible particles from
dark matter annihilation. We have worked out the parameter space and
have found that both the large cross section required for INTEGRAL
and the small cross section for Fermi-LAT can be simultaneously
accommodated in our model. The thermal relic abundance of dark
matter is achieved by the late decay of a charged particle, which
also generates the charge-radius operator. The other charged particle responsible for the charge-radius operator is predicted to have a mass below 1 TeV. The 14 TeV LHC
will concretely test the scenario presented in this paper.

 \subsection*{Acknowledgments}
We would like to thank James Cline, Tim Cohen, Douglas Finkbeiner,
JoAnne Hewett, Dan Hooper, Jessie Shelton, Tracy Slatyer, Aaron
Vincent and Jay Wacker for useful discussions and comments. YB is
supported by start-up funds from the University of Wisconsin,
Madison. YB thanks SLAC for their warm hospitality. SLAC is operated
by Stanford University for the US Department of Energy under
contract DE-AC02-76SF00515.  We also thank the Aspen Center for
Physics, under NSF Grant No. 1066293, where part of this work was
completed. Support for the work of M.S. was provided by NASA through
Einstein Postdoctoral Fellowship grant number PF2-130102 awarded by
the Chandra X-ray Center, which is operated by the Smithsonian
Astrophysical Observatory for NASA under contract NAS8-03060.

\providecommand{\href}[2]{#2}\begingroup\raggedright\endgroup

\end{document}